\documentclass[prl,showpacs,amssymb,floatfix,twocolumn]{revtex4}
\usepackage{amsmath}
\usepackage{graphicx}
\usepackage{epsfig}
\usepackage{dcolumn}
\usepackage{bm}
\usepackage{amssymb,amsmath}

\def\(({\left(}
\def\)){\right)}

\def\[[{\left[}
\def\]]{\right]}

\newcommand{\be}{\begin{equation}}
\newcommand{\ee}{\end{equation}}
\newcommand{\bea}{\begin{eqnarray}}
\newcommand{\eea}{\end{eqnarray}}

\newcommand{\cin}{c_\mathrm{in}}
\newcommand{\cout}{c_\mathrm{out}}

\begin{document}

\title{Phase transition in the detection of modules in sparse
  networks}

\author {Aurelien Decelle$^{1}$, Florent Krzakala$^{2}$, Cristopher Moore$^{3}$, and Lenka Zdeborov\'a$^{4,}$}
\email[Corresponding author; ]{lenka.zdeborova@cea.fr}
\affiliation{
$^1$Universit\'e Paris-Sud \& CNRS, LPTMS, UMR8626,  B\^{a}t.~100, Universit\'e Paris-Sud 91405 Orsay, France.
$^2$CNRS and ESPCI ParisTech, 10 rue Vauquelin, UMR 7083 Gulliver, Paris 75000, France.
$^3$Computer Science Department, University of New Mexico, and the Santa Fe Institute.
$^4$Institut de Physique Th\'eorique, IPhT, CEA Saclay, and URA 2306, CNRS, 91191 Gif-sur-Yvette, France.
}

\begin{abstract}
\vspace{-2mm}
We present an asymptotically exact analysis of the problem of detecting communities in sparse random networks.  
Our results are also applicable to detection of functional modules, partitions, and colorings in noisy planted models.  
Using a cavity method analysis, we unveil a phase transition from a region where the original group assignment is undetectable to one where detection is possible.  In some cases, the detectable region splits into an algorithmically hard region and an easy one.  Our approach naturally translates into a practical algorithm for detecting modules in sparse networks, and learning the parameters of the underlying model.
\end{abstract}

\pacs{\vspace{-1mm} 64.60.aq,89.75.Hc,75.10.Hk}

\maketitle

In many networks, ranging from online communities to food webs, metabolic networks, and genetic regulatory networks, there are communities or modules that play distinct functional roles.  
A fundamental problem is to detect these communities and understand what role they play in the network's structure and dynamics.  
In social networks, these communities are often \emph{assortative}, meaning that there is a higher density of connections within communities than between them, and many approaches to detecting these communities have been proposed (see e.g.~\cite{Fortunato10}).  In other networks, however, these modules may consist of nodes with few connections to each other, but which connect to the rest of the network in similar ways.  

In this Letter we analyze a random generative model for sparse modular networks, known as the stochastic block model.  It provides a useful playground for theoretical ideas and the analysis of algorithms, and is a popular model for functional modules in real networks.  Using the cavity method
developed in the physics of disordered systems~\cite{MezardParisi01,MezardMontanari07} we exactly analyze the detectability of these modules in the limit of large sparse networks.  As a function of the parameters, we compute the phase diagram and locate the associated phase transitions.  

We distinguish between a \emph{detectable} phase where it is possible to learn the model's parameters and the group assignments of the nodes, and a non-intuitive \emph{undetectable} phase where learning is impossible because the network's topology does not retain enough information about the original group memberships.  
The existence of a phase where a certain class of algorithms is unable to detect communities was previously predicted~\cite{ReichardtLeone08}, but its location was only found approximately (and its size overestimated).  
In addition, unlike previous works based on finding a ground state, i.e., minimizing a cost function associated with a group assignment~\cite{ReichardtLeone08,GoodMontjoye10}, our analysis is more general as it relies on the properties of the entire Boltzmann distribution of group assignments.

We also unveil a transition from an algorithmically ``hard'' phase, where, we believe, no polynomial algorithm for learning the groups and parameters exists, to an ``easy'' phase where polynomial algorithms do exist.  In the latter phase, we show that Belief Propagation (BP)~\cite{YedidiaFreeman03} works on large networks in essentially linear time as a function of their size.  BP was previously proposed for community detection~\cite{Hastings06} without, however, the ability to learn the parameters of the underlying model. 

Our approach also provides a natural measure of the significance of the modules in the network, since it computes the marginal probability that a given node belongs to a given group.  If the network does not contain any modules, our method correctly infers this fact by making these marginals uniform. This is an aspect missing in the vast majority of the present approaches to community detection. Our theoretical understanding and algorithm are also applicable to real world networks, as we discuss briefly at the end of this paper (and in detail elsewhere).  Moreover our approach is not restricted by the details of the generative model, and is easily generalized to more elaborate models (e.g., those of~\cite{KarrerNewman10}).

\paragraph*{Stochastic block models.}  We consider networks of $N$ nodes.  Each node $i$ has a hidden label $t_i \in \{1,\ldots,q\}$, specifying which of $q$ groups it is a member of.  These labels are chosen independently, where $n_a$ is the probability that a given node has label $a \in \{1,\ldots,q\}$ (normalized so that $\sum_{a=1}^q n_a=1$).  If $N_a$ is the number of nodes in each group, we have $n_a = \lim_{N \to \infty} N_a/N$.  

Once the group assignment is chosen, the model generates a graph $G$ as follows. For each pair of nodes $i,j$ with $i<j$, we put an edge between $i$ and $j$ independently with probability $p_{t_i,t_j}$, leaving them unconnected with probability $1-p_{t_i,t_j}$.  We call $p_{ab}$ the \emph{affinity} matrix.  Since we are interested in the sparse case where $p_{ab}=O(1/N)$, we will use the rescaled affinity matrix $c_{ab}=Np_{ab}$ and assume that $c_{ab}=O(1)$ in the limit $N \to \infty$.  

In our setting, the adjacency matrix $A_{ij}$ of the graph is the only information available to us.  Our goal is to learn the parameters $q, \{n_a\}, \{p_{ab} \}$ of the block model, as well as the true group assignments $\{t_i\}$.  Special cases of this model have often been considered in the literature.  Planted partitioning, when $n_a=1/q$, $c_{ab}=\cout$ for $a \neq b$ and $c_{aa}=\cin$ with $\cin > \cout$, is a classical problem in computer science
and has been used as a benchmark for community detection~\cite{NewmanGirvan04,Hastings06,ReichardtLeone08,HofmanWiggins08,Fortunato10}.  Planted coloring, where $n_a=1/q$, $c_{aa}=0$, and $c_{ab}=c q/(q-1)$, is a fundamental problem in constraint optimization~\cite{MezardMontanari07}, and was studied using the cavity method in~\cite{KrzakalaZdeborova09}.

\paragraph*{Bayesian inference for block models.}  
Bayesian inference has been applied to community detection before.  However, except for some very specific generative models~\cite{NewmanLeicht07}, the likelihood function must be computed approximately, either through Monte Carlo sampling (e.g.~\cite{ClausetMoore08}) or variational methods~\cite{HofmanWiggins08}. The crucial contribution of our work is that the quantities that follow from Bayesian inference can be computed \emph{exactly} in the thermodynamic limit using the cavity method, or on real finite networks using the BP algorithm in time roughly linear in the size of the network.  The probability that the model parameters take a given set of values $\{ \theta \} = (q,\{n_a\},\{c_{ab}\})$, conditioned on the topology of the network $G$, is
\be 
P(\{\theta\} \mid G) =
\frac{P(\{\theta\})}{P(G)} \sum_{\{t_i\}} P(G,\{t_i\} \mid \{\theta\}) 
\, . \label{P_theta} 
\ee
The sum is over all possible group assignments $\{t_i\}$, where $t_i \in \{1,\ldots,q\}$ for each node $i$.  The prior $P(\{\theta\})$ includes all graph-independent information about the values of the parameters.  We will assume there is no such information available and hence this prior is uniform.  In that case, maximizing $P(\{\theta\} \mid G)$ over $\{\theta\}$ is equivalent to maximizing the sum $\sum_{\{t_i\}} P(G,\{t_i\} \mid \{\theta\})$.  

The function $P(G,\{t_i\} \mid \{\theta\})$ is called the \emph{likelihood}.  It is the probability that the model would produce the group assignment $\{t_i\}$ and the network $G$, assuming that its parameters are $\{\theta\}$.  We can write the likelihood exactly for many different generative models; for the stochastic block model defined above, it is
 \begin{align*} 
 P(G,\{t_i\} \mid \{\theta\} ) 
= \prod_i n_{t_i} \prod_{i<j} \left[ p_{t_i,t_j}^{A_{ij}} (1-p_{t_i,t_j})^{1-A_{ij}}\right] 
 \, .  
  \end{align*}
Thus $P(\{\theta\} \mid G)$ is proportional to  
the partition sum $Z(\{\theta\})$ of a generalized Potts model, with Hamiltonian
\begin{multline} {\cal H}(\{t_i\} \mid \{\theta\}) = - \sum_i \log{n_{t_i}}
\\ 
 - \sum_{i<j} \left[ {A_{ij} \log{ c_{t_i,t_j}}} +
  {(1-A_{ij}) \log{ \left(1-\frac{c_{t_i,t_j}}{N}\right)}}\right] 
  \, . \label{Ham} 
\end{multline}
There is a strong $O(1)$ interaction between connected nodes, and a weak $O(1/N)$ one between unconnected nodes.  The $\log n_{t_i}$ play the role of local fields, enforcing the prior distribution $\{n_a\}$ on group assignments.  

Inferring the parameters $\{\theta\}$ is equivalent to minimizing the free energy $f(\{\theta\})=-\log{Z(\{\theta\})}/N$ associated with~\eqref{Ham}.  If $f(\{\theta\})$ has a non-degenerate minimum, then, from the saddle point method, $\{\theta\}$ is with high probability exactly the set of parameters used in the generation of the network.  In that case, inferring the parameters of the underlying model is possible.

Assuming that we know, or have learned, the correct parameters $\{\theta\}$, how should we determine the group assignment of the nodes?  The most likely assignment $\{t_i\}$ is the ground state of the Hamiltonian~\eqref{Ham}.  However, if we want to find an assignment $\{t_i\}$ that maximizes the number of correctly labeled nodes, we need to follow a different strategy.  Namely, we should compute the marginal distribution $\nu_i(t_i)=\sum_{\{t_j\}_{j\neq i}} \mu(\{t_j\}_{j\neq i},t_i)$ of the label of each node $i$, where $\mu$ is the Boltzmann distribution of~\eqref{Ham}.  
The most probable group assignment for node $i$ is then $t^*_i=\mathrm{argmax}_{t_i} \nu_i(t_i)$.  

It can be proven in general~\cite{NishimoriBook01} that this marginalization maximizes the number of correctly labeled nodes in the thermodynamic limit, and that it is a better choice than using the ground state of~\eqref{Ham}. Furthermore, a configuration chosen according to the Boltzmann distribution has, asymptotically, the correct group sizes and the correct number of edges between each pair of groups, while for the ground state this is not true; finding the minimum bisection, for instance, creates the illusion of two groups even in a completely random graph~\cite{coppersmith}.  Moreover marginalization is algorithmically easier than searching for the ground state.  The expected number of correctly labeled nodes can be estimated as $\sum_i \nu_i(t_i^*)$, even without knowing the original assignment.

\paragraph*{Belief Propagation.} We could estimate the free energy 
using Monte Carlo (MC) sampling, and we do this for comparison.  But a faster algorithm is Belief Propagation (BP), known in physics as the cavity method~\cite{MezardParisi01,MezardMontanari07}.  It is exact in the thermodynamic limit as long as the network is locally treelike, and as long as connected correlations decay rapidly as a function of topological distance. 

To derive the BP equations~\cite{YedidiaFreeman03,MezardMontanari07}, one introduces ``messages'' $\psi^{i\to j}_{t_i}$ and $\psi^{j\to i}_{t_j}$ for each pair of nodes $(i,j)$.  These are conditional marginals in the cavity method.  For instance, $\psi^{i\to j}_{t_i}$ is the probability that $i$ would be in group $t_i$ if $j$ were removed from the network.  Assuming conditional independence between the neighbors of each node and neglecting lower order terms, the messages must be a fixed point of a consistency equation, 
\be \psi_{t_i}^{i\to j} = \frac{1}{Z^{i\to j}} \, n_{t_i} e^{-h_{t_i}}
\prod_{k\in \partial i\setminus j} \left[ \sum_{t_k} c_{t_k t_i}
  \psi_{t_k}^{k\to i} \right] 
  \label{BP_iter} 
  \ee
for each edge $(i,j)$. Here $\partial i$ is the set of $i$'s neighbors, the field $h_{t_i} = \frac{1}{N}\sum_k \sum_{t_k} c_{t_k t_i} \psi_{t_k}^{k}$ summarizes the influence of the non-edges, and $Z^{i\to j}$ is a normalizing factor. We start with random messages, and iterate~\eqref{BP_iter} until we reach a fixed point.  This typically takes just a few iterations, and each step takes linear, $O(N)$, time.

\begin{figure}[t]
\includegraphics[width=8.cm]{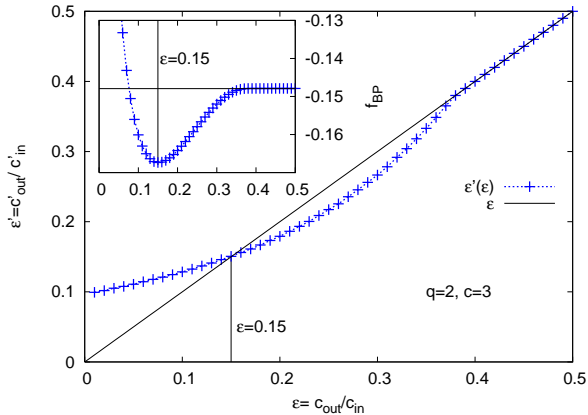}
\caption{Learning for $q=2$ groups with $n_a=1/2$, average degree $c=3$, and $\epsilon=\cout/\cin=0.15$.   If we initialize it in the ordered region, i.e., with $\epsilon_0 < 0.37$, our algorithm infers the correct value of $\epsilon$.  Inset: the free energy as a function of $\epsilon$.  Note the minimum at $\epsilon=0.15$, and the paramagnetic region for $\epsilon > 0.37$.
\label{fig1}}
\end{figure}

The marginals corresponding to the BP fixed point are 
$
\nu_i(t_i) = (1/Z^i) \, 
n_{t_i} e^{-h_{t_i}} \prod_{j\in \partial i} \left[
\sum_{t_j} c_{t_j t_i} \psi_{t_j}^{j\to i} \right] 
$, 
and the free energy is
\begin{equation*}
f_{\rm BP}(\{\theta\}) = - \frac{1}{N} \sum_i \log{Z^i} 
+ \frac{1}{N}\sum_{(i,j) \in E} \log{Z^{ij}} -
\frac{c}{2} \, , 
\end{equation*}
where $Z^{ij} = \sum_{a>b} c_{ab} ( \psi^{i\to j}_a \psi_b^{j\to i}+
\psi^{i\to j}_b \psi_a^{j\to i}) + \sum_a c_{aa} \psi^{i\to j}_a
\psi_a^{j\to i}$.  For more details, see~\cite{YedidiaFreeman03,MezardMontanari07}. Requiring that $f_{\rm BP}(\{\theta\})$ is stationary we update the parameters to their most-likely values given the fixed point
\begin{equation*}
c'_{ab} 
= \sum_{(i,j)\in E} c_{ab} (\psi_a^{i\to j} \psi_b^{j\to i}+\psi_b^{i\to j} \psi_a^{j\to i} )/(Z^{ij} n_a n_b N) \, ,
\end{equation*}
and $n'_a=\sum_i \nu_i(a)/N$. Starting with a suitable initial value $\{\theta_0\}$, we compute $\{\theta'\}$ and iterate until convergence (see Fig.~\ref{fig1}), as in the expectation-maximization algorithm~\cite{DempsterLaird77}. To learn the number of groups $q$, we run the algorithm with several values of $q'$.  The free energy $f_{\rm BP}$ decreases with $q$ and then stays constant for $q' \ge q$.

\paragraph*{Phase diagrams.} 
\begin{figure*}[ht]
\begin{center}
\hspace{-0.6cm}
\includegraphics[width=6.4cm]{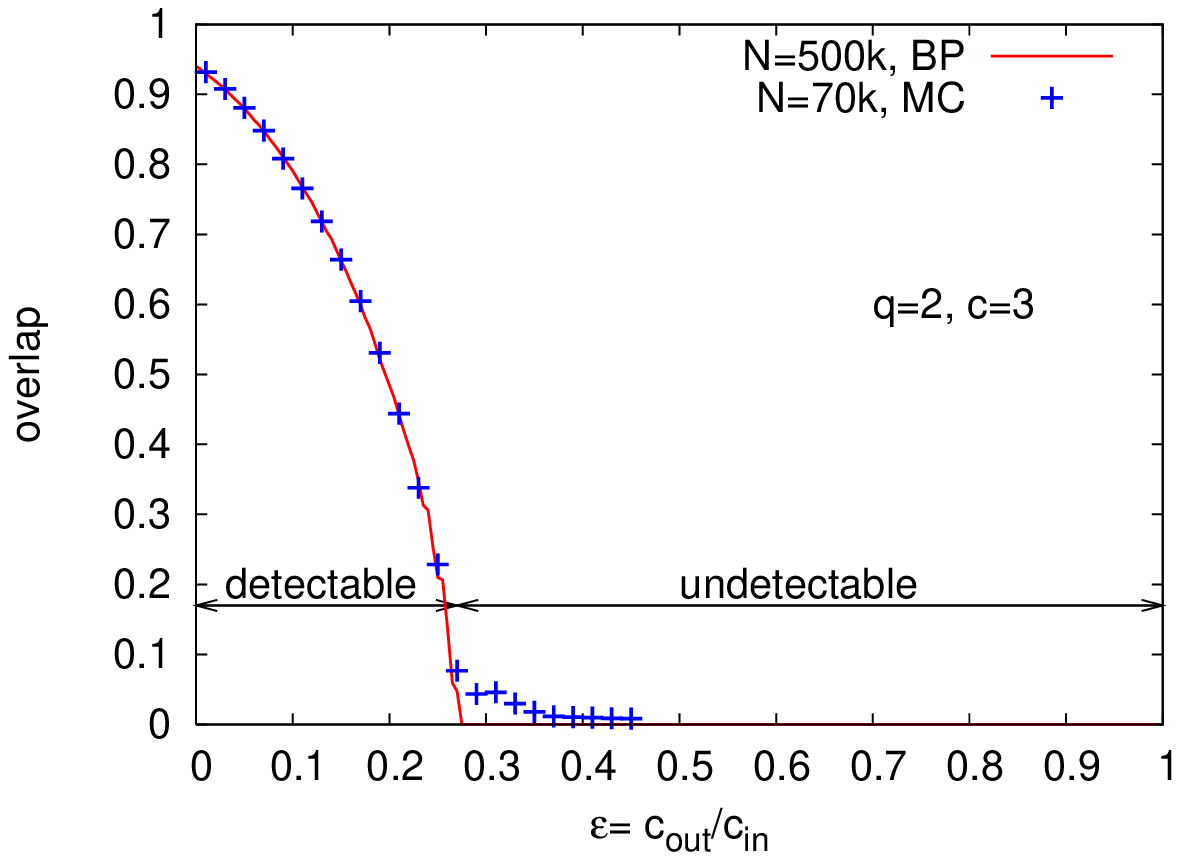}
\hspace{-0.6cm}
\includegraphics[width=6.4cm]{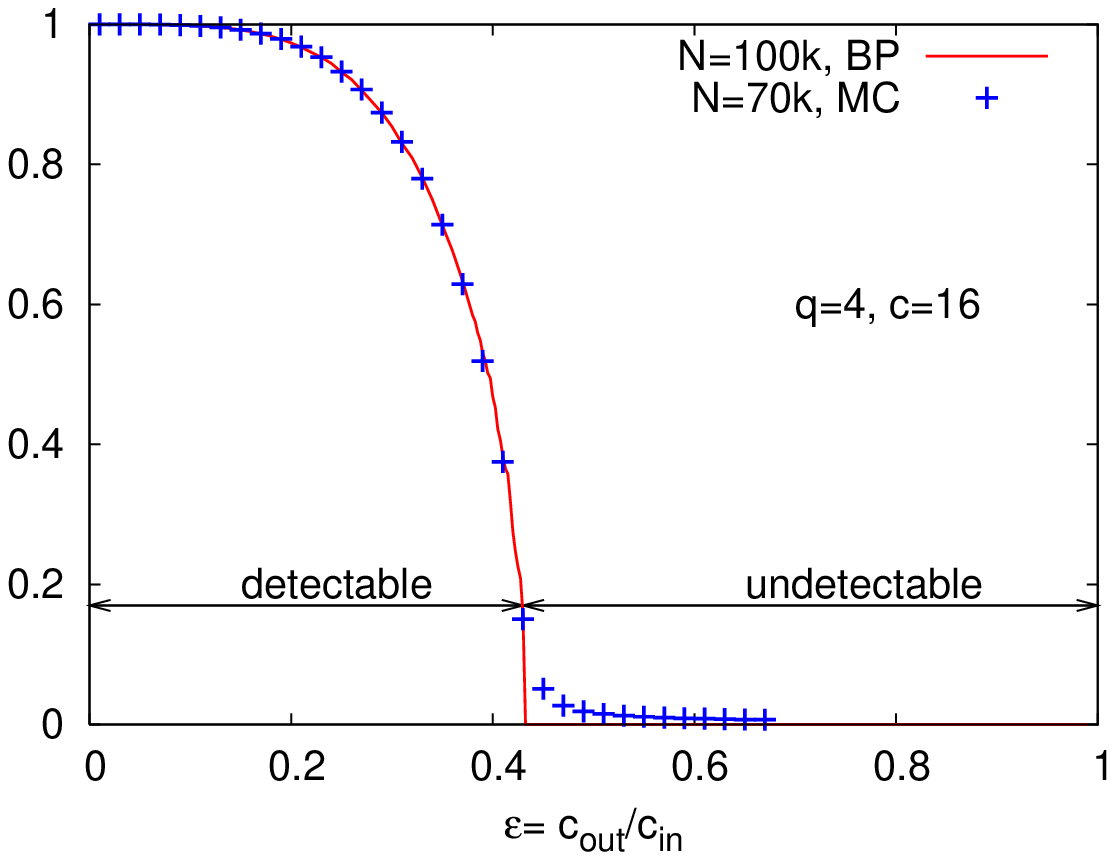}
\hspace{-0.6cm}
\includegraphics[width=6.4cm]{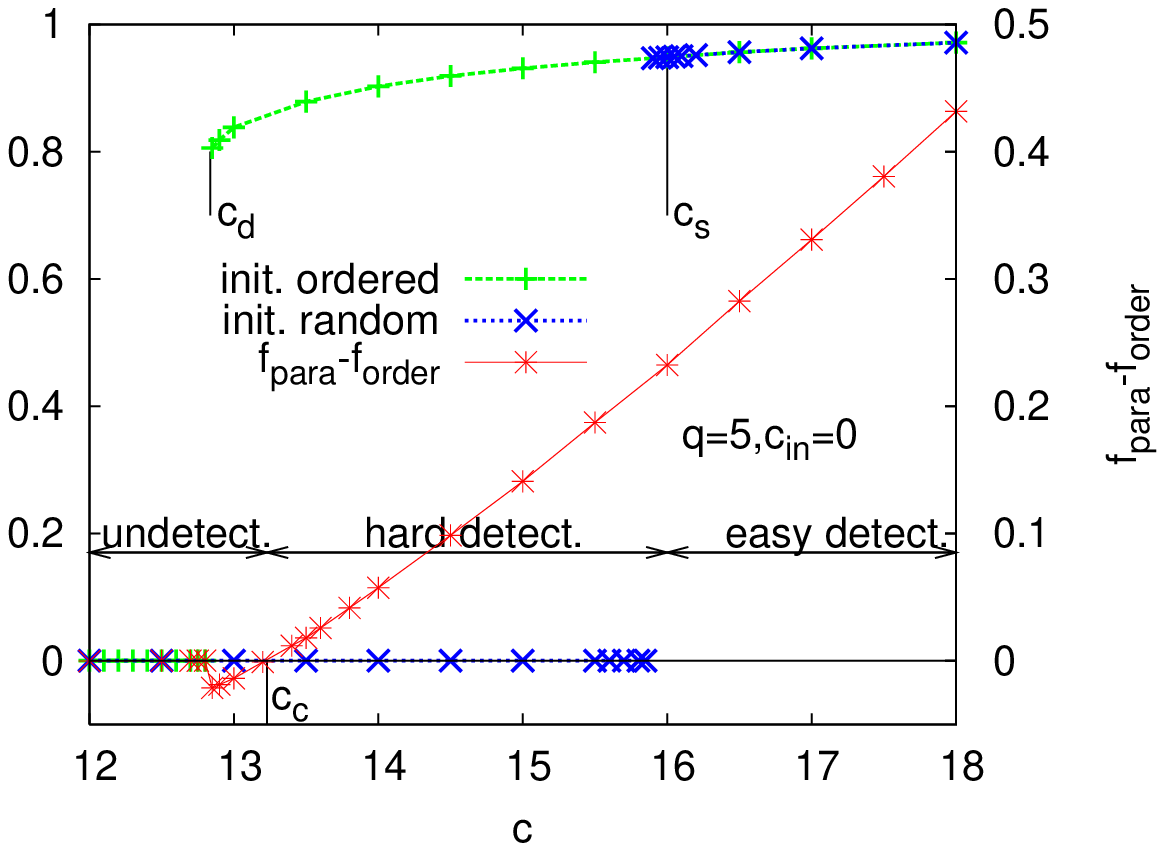}
\vspace{-6mm}
\end{center}
\caption{(color online) The best possible overlap between the inferred and
  original group assignment. Left: community detection with $q=2$,
  $c=3$ and different values of $\epsilon=\cout/\cin$.  A continuous phase transition between a detectable and a
  non-detectable phase arises at the critical point given by~\eqref{stab}.  Middle: The 4-group community detection benchmark
  of~\cite{NewmanGirvan04} with $c=16$, with the same
  phenomenology. The results agree well with MC simulations,
  except very close to the critical point where finite-size effects
  are stronger.  Right: A planted coloring problem with $q=5$ and
  $\cin=0$, $c=\cout(1-1/q)$. Both the ordered fixed point (green $+$s, obtained by initializing in the actual group assignment)
  and the paramagnetic one (blue $\times$s, obtained by initializing the algorithm in a random configuration) exist between $c_d$ and $c_s$. 
The difference $\Delta f$ (red) between the 
  paramagnetic and ordered free energies shows that modules are in
  principle detectable as soon as $c>c_c$ when $\Delta f>0$. It is in
  practice impossible to find the corresponding fixed point, and
  detection become feasible only after the spinodal point $c_s$ given
  by~\eqref{stab}.\label{fig2} \vspace{-2mm} }
\end{figure*}

For illustration we use the case of planted partitions and colorings, $n_a=1/q$, $c_{ab}=\cout$ for $a \neq b$, and $c_{aa}=\cin$.  We observe three different cases governing the free energy landscape $f_{\rm BP}{{\{\theta}\}}$.  In the ``paramagnetic'' phase, the free energy is constant in the vicinity of the true value of $\{\theta\}$.  Learning is impossible, and the marginals are $\nu_i(t_i)=1/q$ for all nodes.  In this case the overlap between the original assignment and the one resulting from BP marginalization, defined as
 \be
 Q(\{t_i\},\{q_i\}) = \max_{\pi \in S_q}\frac{
  (1/N) \, \sum_{i} \delta_{t_i,\pi(q_i)} - \max_a n_a
}{1-\max_a n_a} \, , \label{overlap} 
\ee
%
(where $S_q$ is the permutation group) is zero, and the original assignment is undetectable.  Generalizing~\cite{AchlioptasCoja-Oghlan08-focs,KrzakalaZdeborova09}, one can show there is essentially no difference between a graph produced by the block model and a completely random graph of the same average degree; the free energy of the two ensembles is asymptotically identical. 

In the \emph{ordered} phase, $f_{\rm BP}$ has an attractive global minimum at the true value of $\{\theta\}$, and BP rapidly infers the correct parameters.  
This is illustrated in Fig.~\ref{fig2}.  As $\epsilon=\cout/\cin$ varies from $0$ ($q$ completely separated groups) to $1$ (a pure random graph), we observe a continuous phase transition from an ordered phase with positive overlap to a paramagnetic phase with zero overlap.  Thus there is a second-order transition from a detectable to an undetectable phase.

A third situation arises if $f_{\rm BP}{{\{\theta}\}}$ has both a paramagnetic fixed point \emph{and} the ordered fixed point at the true~$\{\theta\}$.  In this case, the two phases co-exist and the detectability transition is first-order; see Fig.~\ref{fig2} on the right. The phase transition is located by comparing the free energies of the two phases. However, even if the ordered fixed point has a lower free energy, it is not easy to find it unless the initial messages are close to the true group assignment. All but an exponentially small set of initial messages will lead to the paramagnetic fixed point.  
This situation is typical of mean-field first-order phase transitions. In fact, recent results about random optimization problems show that finding the lower-free-energy phase in this case is an extremely hard problem~\cite{FranzMezard01,KrzakalaZdeborova09}. 

Only when the paramagnetic phase is no longer locally stable does inference become easy.  We can compute the location of the transition to this easily-detectable phase analytically by analyzing how a small random perturbation to the paramagnetic fixed point propagates as the BP equations are iterated~\cite{MezardMontanari06,KrzakalaZdeborova09}.  It follows that for
\be 
|\cin-\cout| > q \sqrt{c} \label{stab} \, , 
\ee
the original group assignment is dynamically attractive and hence many algorithms, e.g. MC or BP, will converge to it.  
Note that it is typically still hard to compute the 
ground state of~\eqref{Ham}, even though we can compute the marginals, and therefore the optimal estimate of the group assignment, asymptotically exactly.

On the other hand, if~\eqref{stab} is not satisfied then community detection is either impossible, or at best as hard as solving the hardest known optimization problems.  When $\cout < \cin$ the phase transition is of first-order for $q>4$, as can be retrieved from data presented in~\cite{MezardMontanari06}.  However, the detectable but hard region is so narrow that is is quite unlikely to appear in realistic situations. 

\paragraph*{Real-world networks.}  
Our algorithm is not restricted to large random networks; it is applicable to real networks as well.  We tested it on the ``Karate Club'' network~\cite{Zachary77}, a common benchmark for community detection.  For $q=2$, BP leads to two different fixed points.  One corresponds to the actual known division into two groups.  
The other has a smaller free energy and thus a larger likelihood, 
and splits the network into high-degree nodes and low-degree nodes as found in~\cite{KarrerNewman10}.  These two fixed points correspond to two local minima of $f_{\rm BP}$ for $q=2$, and depending on the initial value $\{\theta_0\}$ 
BP converges to one or the other.  
For $q>2$, our algorithm converges to fixed points with yet lower values of $f_{\rm BP}$. 
For $q=4$ the best fixed point corresponds to a splitting of the two actual groups into high-degree and low-degree subgroups.  

The results obtained with MC, which can be easily equilibrated for such a small network, are almost identical to those of BP in terms of the parameters and marginals, and identical in terms of the estimated group assignments.  This demonstrates that BP is a useful approach even on real, finite networks that are far from trees.

\paragraph*{Conclusion.}  
We have presented a principled and asymptotically exact analysis of the detection of communities in networks generated by the stochastic block model.  There is a strict limit on detectability due to a transition from a phase where the free energy landscape lets us infer the model's parameters, to a phase where it does not.  In some cases the communities are detectable, but the problem is hard because the attractive region around the correct fixed point is exponentially small.  Our analysis comes with an associated learning algorithm, which for large sparse networks generated from the model is able to learn the number of groups, their exact sizes, and the affinity matrix $p_{ab}$.  Our approach and our algorithm are easily generalized to other local generative models, and we will investigate its performance on a variety of real-world networks in the future.

\paragraph*{Acknowledgments.} We are grateful to Mark Newman for helpful discussions. C.M. is funded by the McDonnell Foundation.


\bibliographystyle{prsty}
\bibliography{myentries}

\end{document}